\documentclass[letterpaper,11pt]{article}

\usepackage{todonotes}
\usepackage{array,multirow, graphicx}
\usepackage{enumitem}
\usepackage{float}
\usepackage[margin=1in]{geometry}

\usepackage{makecell}
\usepackage{colortbl}
\usepackage{hhline}
\usepackage{quoting}
\usepackage{cite}
\usepackage{tablefootnote}

\usepackage[hyphens,spaces,obeyspaces]{url}
\usepackage{hyperref}
\usepackage[hyphenbreaks]{breakurl}

\usepackage[hang,flushmargin]{footmisc}

\newcommand{\Csharp}{%
  {\settoheight{\dimen0}{C}C\kern-.05em \resizebox{!}{\dimen0}{\raisebox{\depth}{\#}}}}

\def\BibTeX{{\rm B\kern-.05em{\sc i\kern-.025em b}\kern-.08emT\kern-.1667em\lower.7ex\hbox{E}\kern-.125emX}}

\title{\vspace{-3em}{\Huge Cloud Programming Simplified:\\A Berkeley View on Serverless Computing}}

\author{
\protect{\begin{tabular}{cccc}
  Eric Jonas & Johann Schleier-Smith & Vikram Sreekanti & Chia-Che Tsai\\
  Anurag Khandelwal & Qifan Pu & Vaishaal Shankar & Joao Carreira\\
  Karl Krauth &Neeraja Yadwadkar & Joseph E. Gonzalez & Raluca Ada Popa\\
  & Ion Stoica & David A. Patterson & \\
\end{tabular}}
}

\date{%
    UC Berkeley\\
    \vspace{8pt}
    \texttt{serverlessview@berkeley.edu}\\%
}

\begin{document}

\maketitle


%

\renewenvironment{abstract}
 {\small
  \begin{center}
  \bfseries \abstractname\vspace{-.5em}\vspace{0pt}
  \end{center}
  \list{}{
    \setlength{\leftmargin}{.1cm}%
    \setlength{\rightmargin}{\leftmargin}%
  }%
  \item\relax}
 {\endlist}
\begin{abstract}
Serverless cloud computing handles virtually all the system administration operations needed to make it easier for programmers to use the cloud. It provides an interface that greatly simplifies cloud programming, and represents an evolution that parallels the transition from assembly language to high-level programming languages. This paper gives a quick history of cloud computing, including an accounting of the predictions of the 2009 Berkeley View of Cloud Computing paper, explains the motivation for serverless computing, describes applications that stretch the current limits of serverless, and then lists obstacles and research opportunities required for serverless computing to fulfill its full potential. Just as the 2009 paper identified challenges for the cloud and predicted they would be addressed and that cloud use would accelerate, we predict these issues are solvable and that serverless computing will grow to dominate the future of cloud computing.
\end{abstract}

\tableofcontents

\pagebreak
\section{Introduction to Serverless Computing}
\label{sec:sec1}
\vspace{\baselineskip}

\begin{quoting}
\emph{The data center is now the computer.}
\begin{flushright}
Luiz Barroso (2007)~\cite{patterson2008data}
\end{flushright}
\end{quoting}
\vspace{\baselineskip}
In 2009, to help explain the excitement around cloud computing, ``The Berkeley View on Cloud Computing''~\cite{Armbrust09abovethe} identified six potential advantages:
\begin{enumerate}[noitemsep]
    \item The appearance of infinite computing resources on demand.
    \item The elimination of an up-front commitment by cloud users. 
    \item The ability to pay for use of computing resources on a short-term basis as needed. 
    \item Economies of scale that significantly reduced cost due to many, very large data centers. 
    \item Simplifying operation and increasing utilization via resource virtualization.
    \item Higher hardware utilization by multiplexing workloads from different organizations.
\end{enumerate}

The past ten years have seen these advantages largely realized, but cloud users continue to bear a burden from complex operations and many workloads still do not benefit from efficient multiplexing. These shortfalls mostly correspond to failures to realize the last two potential advantages. Cloud computing relieved users of physical infrastructure management but left them with a proliferation of virtual resources to manage. Multiplexing worked well for batch style workloads such as MapReduce or high performance computing, which could fully utilize the instances they allocated. It worked less well for stateful services, such as when porting enterprise software like a database management system to the cloud.\footnote{Due to the tight coupling between computation and storage, databases need to reserve instances long term. However, their workloads can be bursty, which results in low resource utilization.}

In 2009, there were two competing approaches to virtualization in the cloud. As the paper explained:
\begin{center}
\begin{minipage}{.9\linewidth}
\emph{Amazon EC2 is at one end of the spectrum. An EC2 instance looks much like physical hardware, and users can control nearly the entire software stack, from the kernel upward. ... At the other extreme of the spectrum are application domain-specific platforms such as Google App Engine ... enforcing an application structure of clean separation between a stateless computation tier and a stateful storage tier. App Engine's impressive automatic scaling and high-availability mechanisms ... rely on these constraints.}
\end{minipage}
\end{center}
The marketplace eventually embraced Amazon's low-level virtual machine approach to cloud computing, so Google, Microsoft and other cloud companies offered similar interfaces. We believe the main reason for the success of low-level virtual machines was that in the early days of cloud computing users wanted to recreate the same computing environment in the cloud that they had on their local computers to simplify porting their workloads to the cloud~\cite{orban2017aws,Delimitrou:quasar,mars2011bubble,Yang:bubbleflux}. That practical need, sensibly enough, took priority over writing new programs solely for the cloud, especially as it was unclear how successful the cloud would be.

The downside of this choice was that developers had to manage virtual machines themselves, basically either by becoming system administrators or by working with them to set up environments. Table~\ref{tbl:vm_issues} lists the issues that must be managed to operate an environment in the cloud. The long list of low-level virtual machine management responsibilities inspired customers with simpler applications to ask for an easier path to the cloud for new applications. For example, suppose the application wanted to send images from a phone application to the cloud, which should create thumbnail images and then place them on the web.  The code to accomplish these tasks might be dozens of lines of JavaScript, which would be a trivial amount of development compared to what it takes to set up the servers with the proper environment to run the code. 

Recognition of these needs led to a new option from Amazon in 2015 called the AWS Lambda service.  Lambda offered \textit{cloud functions}, and drew widespread attention to \textit{serverless computing}. Although “serverless computing” is arguably an oxymoron\textemdash{}you are still using servers to compute\textemdash{}the name presumably stuck because it suggests that the cloud user simply writes the code and leaves all the server provisioning and administration tasks to the cloud provider. While cloud functions\textemdash{}packaged as FaaS (Function as a Service) offerings\footnote{Different cloud platforms have different names for their FaaS offerings\textemdash{}AWS Lambda for Amazon Web Services (AWS), Google Cloud Functions for Google Cloud Platform, IBM Cloud Functions for IBM Cloud, and Azure Functions for Microsoft Azure. They all have similar features, and we refer to them as cloud functions or FaaS offerings interchangeably in this paper.}\textemdash{}represent the core of serverless computing, cloud platforms also provide specialized serverless frameworks that cater to specific application requirements as BaaS (Backend as a Service) offerings~\cite{serverless_computing}. Put simply, \textit{serverless computing = FaaS + BaaS}.\footnote{BaaS originated as a term describing mobile-centric cloud frameworks and has grown to encompass any application-specific serverless cloud service, such as serverless databases and serverless big data processing frameworks.} In our definition, for a service to be considered serverless, it must scale automatically with no need for explicit provisioning, and be billed based on usage. In the rest of this paper, we focus on the emergence, evolution, and future of cloud functions. Cloud functions are the \textit{general purpose} element in serverless computing today, and lead the way to a simplified and general purpose programming model for the cloud.

We next motivate and define serverless computing. Like the original Berkeley View paper on cloud computing, we then list challenges and research opportunities to be addressed for serverless computing  to fulfill its promise. While we are unsure which solutions will win, we believe all issues will all be addressed eventually, thereby enabling serverless computing to become the face of cloud computing.

\begin{table}
\fbox{\parbox{\textwidth}{
\begin{enumerate}[noitemsep]
    \item Redundancy for availability,  so that a single machine failure doesn't take down the service.
    \item Geographic distribution of redundant copies to preserve the service in case of disaster.
    \item Load balancing and request routing to efficiently utilize resources.
    \item Autoscaling in response to changes in load to scale up or down the system.
    \item Monitoring to make sure the service is still running well.
    \item Logging to record messages needed for debugging or performance tuning.
    \item System upgrades, including security patching.
    \item Migration to new instances as they become available.
\end{enumerate}
}}
\label{tbl:vm_issues}
\caption{Eight issues to be addressed in setting up an environment for cloud users. Some issues take many steps. For example, autoscaling requires determining the need to scale; picking the type and number of servers to use; requesting the servers; waiting for them to come online; configuring them with the application; confirming that no errors occurred; instrumenting them with monitoring tools; and sending traffic at them to test them.}
\end{table}

\section{Emergence of Serverless Computing}
\label{sec:sec2}

In any serverless platform, the user just writes a cloud function in a high-level language, picks the event that should trigger the running of the function\textemdash{}such as loading an image into cloud storage or adding an image thumbnail to a database table\textemdash{}and lets the serverless system handle everything else: instance selection, scaling, deployment, fault tolerance, monitoring, logging, security patches, and so on. Table~\ref{tbl:serverless_vs_serverful} summarizes the differences between serverless and the traditional approach, which we'll call \textit{serverful cloud computing} in this paper. Note that these two approaches represent the endpoints of a continuum of function-based/server-centered computing platforms, with containerized orchestration frameworks like Kubernetes representing intermediates.

\renewcommand{\arraystretch}{1.2} 

\begin{table}[h]
\centering
\resizebox{\columnwidth}{!}{%
\begin{tabular}{|l|l|l|l|}
 \hline 
                            & \multicolumn{1}{c|}{\textit{Characteristic}}            & \multicolumn{1}{c|}{\textit{AWS Serverless Cloud}}                    & \multicolumn{1}{c|}{\textit{AWS Serverful Cloud}}                   \\ \hline 
\parbox[t]{2mm}{\multirow{8}{*}{\rotatebox[origin=c]{90}{PROGRAMMER}}} & When the program is run   & On event selected by Cloud user         & Continuously until explicitly stopped \\ \cline{2-4} 
                            & Programming Language      & JavaScript, Python, Java, Go, \Csharp, etc.\footnote{Note that the number of supported languages has been growing and that AWS has also introduced support for user-contributed language runtimes. Benefiting fully from simplified cloud programming requires using a high-level language, and while users have always been able to run can technically run an arbitrary x86 executable by calling it from a script written in an officially supported programming language, but this reduces the potential of serverless computing because it forces the cloud provider to use an x86 computer instead of enabling more innovative hardware that could improve price-performance (see Section~\ref{sec:sec4}).} & Any                                   \\ \cline{2-4} 
                            & Program State             & Kept in storage (stateless)             & Anywhere (stateful or stateless)      \\ \cline{2-4} 
                            & Maximum Memory Size       & 0.125 - 3 GiB (Cloud user selects)      & 0.5 - 1952 GiB (Cloud user selects)   \\ \cline{2-4} 
                            & Maximum Local Storage     & 0.5 GiB                                 & 0 - 3600 GiB (Cloud user selects)     \\ \cline{2-4} 
                            & Maximum Run Time          & 900 seconds                             & None                                  \\ \cline{2-4} 
                            & Minimum Accounting Unit   & 0.1 seconds                             & 60 seconds                            \\ \cline{2-4} 
                            & Price per Accounting Unit & \$0.0000002 (assuming 0.125 GiB)        & \$0.0000867 - \$0.4080000             \\ \cline{2-4}
                            & Operating System \& Libraries & Cloud provider selects\footnote{Cloud users can add libraries to the default ones provided by the cloud provider.}              & Cloud user selects                    \\ \hline
                            
\parbox[t]{2mm}{\multirow{6}{*}{\rotatebox[origin=c]{90}{SYSADMIN}}} & Server Instance & Cloud provider selects & Cloud user selects      \\ \cline{2-4} 
                            & Scaling\footnote{Autoscaling cannot be quantified single metric. Users may be interested in scaling up the number of concurrent requests or reducing the interval till the first instruction is run. Structuring these guarantees will be up to the provider.}                   & Cloud provider responsible              & Cloud user responsible                \\ \cline{2-4} 
                            & Deployment                & Cloud provider responsible              & Cloud user responsible                \\ \cline{2-4} 
                            & Fault Tolerance           & Cloud provider responsible              & Cloud user responsible                \\ \cline{2-4} 
                            & Monitoring                & Cloud provider responsible              & Cloud user responsible                \\ \cline{2-4} 
                            & Logging                   & Cloud provider responsible              & Cloud user responsible                \\ \hline
\end{tabular}
}
\caption{Characteristics of serverless cloud functions vs. serverful cloud VMs  divided into programming and system administration categories. Specifications and prices correspond to AWS Lambda and to on-demand AWS EC2 instances.}
\label{tbl:serverless_vs_serverful}
\end{table}

\begin{figure}[h]
  \centering
  \includegraphics[width=\columnwidth]{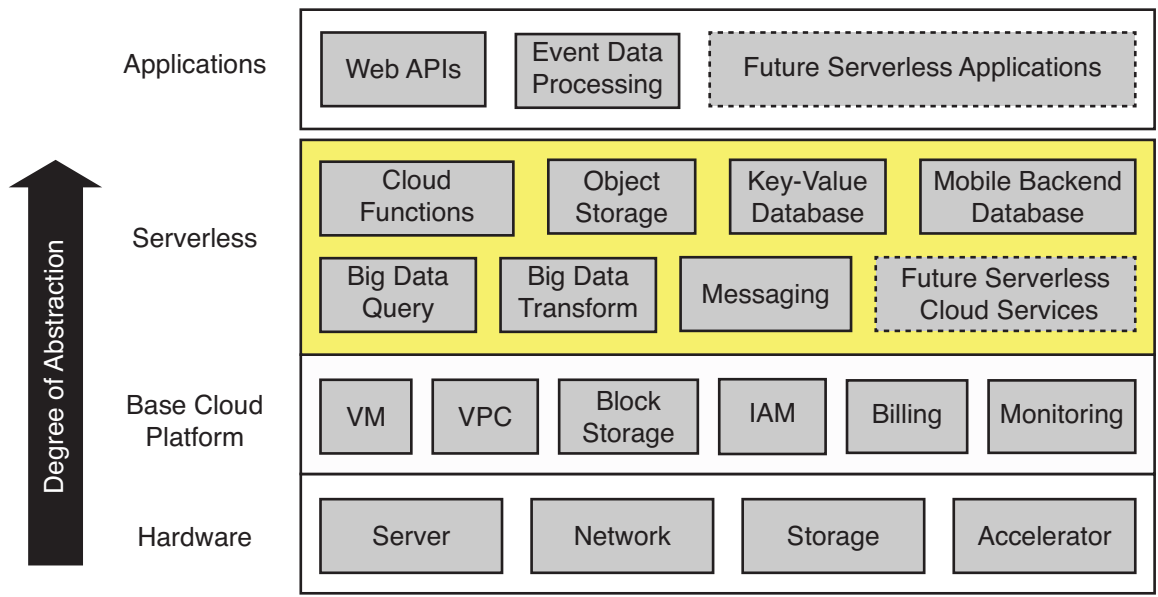}
  \caption{Architecture of the serverless cloud. The serverless layer sits between applications and the base cloud platform, simplifying cloud programming. Cloud functions (i.e., FaaS) provide general compute and are complemented by an ecosystem of specialized Backend as a Service (BaaS) offerings such as object storage, databases, or messaging. Specifically, a serverless application on AWS might use Lambda with S3 (object storage) and DynamoDB (key-value database), while an application on Google's cloud might use Cloud Functions with Cloud Firestore (mobile backend database) and Cloud Pub/Sub (messaging). Serverless also comprises certain big data services such as AWS Athena and Google BigQuery (big data query), and Google Cloud Dataflow and AWS Glue (big data transform). The base underlying base cloud platform includes virtual machines (VM), private networks (VPC), virtualized block storage, Identity and Access Management (IAM), as well as billing and monitoring.}
  \label{fig:serverless_stack}
\end{figure}

Figure~\ref{fig:serverless_stack} illustrates how serverless simplifies application development by making cloud resources easier to use. In the cloud context, serverful computing is like programming in low-level assembly language whereas serverless computing is like programming in a higher-level language such as Python. An assembly language programmer computing a simple expression such as \texttt{c = a + b} must select one or more registers to use, load the values into those registers, perform the arithmetic, and then store the result. This mirrors several of the steps of serverful cloud programming, where one first provisions resources or identifies available ones, then loads those resources with necessary code and data, performs the computation, returns or stores the results, and eventually manages resource release. The aim and opportunity in serverless computing is to give cloud programmers benefits similar to those in the transition to high-level programming languages.\footnote{Although several serverless computing providers run binary programs in addition to high-level language programs, we believe the greatest upside potential for serverless is using high-level languages.} Other features of high-level programming environments have natural parallels in serverless computing as well. Automated memory management relieves programmers from managing memory resources, whereas serverless computing relieves programmers from managing server resources. 

Put precisely, there are three critical distinctions between serverless and serverful computing:

\begin{enumerate}[noitemsep]
    \item \textit{Decoupled computation and storage}. The storage and computation scale separately and are provisioned and priced independently. In general, the storage is provided by a separate cloud service and the computation is stateless. 
    \item \textit{Executing code without managing resource allocation}. Instead of requesting resources, the user provides a piece of code and the cloud automatically provisions resources to execute that code. 
    \item \textit{Paying in proportion to resources used instead of for resources allocated}. Billing is by some dimension associated with the execution, such as execution time, rather than by a dimension of the base cloud platform, such as size and number of VMs allocated.
\end{enumerate}

Using these distinctions, we next explain how serverless differs from similar offerings, both past and current.

\subsection{Contextualizing Serverless Computing}

What technical breakthroughs were needed to make serverless computing possible? Some have argued that serverless computing is merely a rebranding of preceding offerings, perhaps a modest generalization of Platform as a Service (PaaS) cloud products such as Heroku~\cite{heroku}, Firebase~\cite{firebase}, or Parse~\cite{parse}. Others might point out that the shared web hosting environments popular in the 1990s provided much of what serverless computing has to offer. For example, these had a stateless programming model allowing high levels of multi-tenancy, elastic response to variable demand, and a standardized function invocation API, the Common Gateway Interface (CGI)~\cite{cgi}, which even allowed direct deployment of source code written in high-level languages such as Perl or PHP. Google's original App Engine, largely rebuffed by the market just a few years before serverless computing gained in popularity, also allowed developers to deploy code while leaving most aspects of operations to the cloud provider. We believe serverless computing represents significant innovation over PaaS and other previous models.

Today's serverless computing with cloud functions differs from its predecessors in several essential ways: better autoscaling, strong isolation, platform flexibility, and service ecosystem support. Among these factors, the autoscaling offered by AWS Lambda marked a striking departure from what came before. It tracked load with much greater fidelity than serverful autoscaling techniques, responding quickly to scale up when needed and scaling all the way down to zero resources, and zero cost, in the absence of demand. It charged in a much more fine-grained way, providing a minimum billing increment of 100 ms at a time when other autoscaling services charged by the hour.\footnote{Compare for example, AWS Elastic Beanstalk or Google App Engine.} In a critical departure, it charged the customer for the \textit{time their code was actually executing, not for the resources reserved to execute their program}. This distinction ensured the cloud provider had ``skin in the game'' on autoscaling, and consequently provided incentives to ensure efficient resource allocation.

Serverless computing relies on strong performance and security isolation to make multi-tenant hardware sharing possible. VM-like isolation is the current standard for multi-tenant hardware sharing for cloud functions~\cite{wang2018peeking}, but because VM provisioning can take many seconds serverless computing providers use elaborate techniques to speed up the creation of function execution environments. One approach, reflected in AWS Lambda, is maintaining a ``warm pool'' of VM instances that need only be assigned to a tenant, and an ``active pool'' of instances that have been used to run a function before and are maintained to serve future invocations~\cite{wagnercompute}. The resource lifecycle management and multi-tenant bin packing necessary to achieve high utilization are key technical enablers of serverless computing. We note that several recent proposals aim to reduce the overhead of providing multi-tenant isolation by leveraging containers, unikernels, library OSes, or language VMs. For example, Google has announced that gVisor~\cite{gVisor} has already been adopted by App Engine, Cloud Functions, and Cloud ML Engine, Amazon released Firecracker VMs~\cite{firecracker} for AWS Lambda and AWS Fargate, and the CloudFlare Workers serverless platform provides multi-tenant isolation between JavaScript cloud functions using web browser sandboxing technology~\cite{without_containers}.

Several other distinctions have helped serverless computing succeed. By allowing users to bring their own libraries, serverless computing can support a much broader range of applications than PaaS services which are tied closely to particular use cases. Serverless computing runs in modern data centers and operates at much greater scale than the old shared web hosting environments.

\begin{table}[h]
\centering
\resizebox{\columnwidth}{!}{%
\begin{tabular}{|l|l|l|}
\hline
\textit{Service}         & \textit{Programming Interface}  & \textit{Cost Model}                              \\ \hline
Cloud Functions & Arbitrary code         & Function execution time                 \\ \hline
BigQuery/Athena & SQL-like query          & The amount of data scanned by the query \\ \hline
DynamoDB        & puts() and gets()      & Per put() or get() request + storage    \\ \hline
SQS             & enqueue/dequeue events & per-API call                            \\ \hline
\end{tabular}%
}
\caption{Examples of serverless computing services and their corresponding programming interfaces and cost models. Note that for the serverless compute offerings described here: BigQuery, Athena, and cloud functions,  the user pays separately for storage (e.g., in Google Cloud Storage, AWS S3, or Azure Blob Storage).}
\label{tbl:services}
\end{table}

As mentioned in Section~\ref{sec:sec1}, cloud functions (i.e., FaaS) popularized the serverless paradigm. However, it is worth acknowledging that they owe their success in part to BaaS offerings that have existed since the beginning of public clouds, services like AWS S3. In our view, these services are domain-specific, highly optimized implementations of serverless computing. Cloud functions represent serverless computing in a more general form. We summarize this view in Table~\ref{tbl:services} by comparing programming interfaces and cost models for several services.

A common question when discussing serverless computing is how it relates to Kubernetes~\cite{kubernetes}, a ``container orchestration'' technology for deploying microservices. Unlike serverless computing, \textit{Kubernetes is a technology that simplifies management of serverful computing}. Derived from years of development for Google's internal use~\cite{burns2016borg}, it is gaining rapid adoption. Kubernetes can provide short-lived computing environments, like serverless computing, and has far fewer limitations, e.g., on hardware resources, execution time, and network communication. It can also deploy software originally developed for on-premise use completely on the public cloud with little modification. Serverless computing, on the other hand, introduces a paradigm shift that allows fully offloading operational responsibilities to the provider, and makes possible fine-grained multi-tenant multiplexing. Hosted Kubernetes offerings, such as the Google Kubernetes Engine (GKE) and AWS Elastic Kubernetes Service (EKS) offer a middle ground in this continuum: they offload operational management of Kubernetes while giving developers the flexibility to configure arbitrary containers. One key difference between hosted Kubernetes services and serverless computing is the billing model. The former charges per reserved resources, whereas the latter per function execution duration.

Kubernetes is also a perfect match to hybrid applications where a portion runs on-premise on local hardware and a portion runs in the cloud. Our view is that such hybrid applications make sense in the transition to the cloud. In the long term, however, we believe the economies of cloud scale, faster network bandwidth, increasing cloud services, and simplification of cloud management via serverless computing will diminish the importance of such hybrid applications. 

Edge computing is the partner of cloud computing in the PostPC Era, and while we focus here on how serverless computing will transform programming within the data center, there is interesting potential for impact at the edge as well. Several Content Delivery Network (CDN) operators offers the ability to execute a serverless functions in facilities close to users~\cite{lambda_edge,cloudfare_workers}, wherever they might be, and AWS IoT Greengrass~\cite{greengrass} can even embed serverless execution in edge devices.

Now that we've defined and contextualized serverless computing, let's see why it is attractive to cloud providers, cloud users, and researchers.

\subsection{Attractiveness of Serverless Computing}

For cloud providers serverless computing promotes business growth, as making the cloud easier to program helps draw in new customers and helps existing customers make more use of cloud offerings. For example, recent surveys found that about 24\% of serverless users were new to cloud computing~\cite{serverless_growth} and 30\% of existing serverful cloud customers also used serverless computing~\cite{cloud_native}.  In addition, the short run time, small memory footprint, and stateless nature improve statistical multiplexing by making it easier for cloud providers to find unused resources on which to run these tasks. The cloud providers can also utilize less popular computers\textemdash{}as the instance type is up to the cloud providers\textemdash{}such as older servers that may be less attractive to serverful cloud customers. Both benefits increase income from existing resources.

Customers benefit from increased programming productivity, and in many scenarios can enjoy cost savings as well, a consequence of the higher utilization of underlying servers. Even if serverless computing lets customers be more efficient, the Jevons paradox~\cite{jevons_paradox} suggests that they will increase their use of the cloud rather than cut back as the greater efficiency will increase the demand by adding users. 

Serverless also raises the cloud deployment level from x86 machine code\textemdash{}99\% of cloud computers use the x86 instruction set\textemdash{}to high-level programming languages,\footnote{Although several serverless computing providers run binary programs in addition to high-level language programs, we believe the greatest upside potential for serverless is using high-level languages.} which enables architectural innovations. If ARM or RISC-V offer better cost-performance than x86, serverless computing makes it easier to change instruction sets. Cloud providers could even embrace research in language oriented optimizations and domain specific architectures specifically aimed at accelerating programs written in languages like Python~\cite{Hennessy:2019:NGA:3310134.3282307} (see Section~\ref{sec:sec4}).

Cloud users like serverless computing because novices can deploy functions without any understanding of the cloud infrastructure and because experts save development time and stay focused on problems unique to their application. Serverless users may save money since the functions are executed only when events occur, and fine-grained accounting (today typically 100 milliseconds) means they pay only for what they use versus for what they reserve. Table~\ref{tbl:serverless_popularity} shows the most popular uses of serverless computing today.

Researchers have been attracted to serverless computing, and especially to cloud functions, because it is a new general purpose compute abstraction that promises to become the future of cloud computing, and because there are many opportunities for boosting the current performance and overcoming its current limitations.

\begin{table}[]
\centering
\resizebox{\columnwidth}{!}{%
\begin{tabular}{|l|l|}
\hline
\textit{Percent} & \textit{Use Case}                                                                 \\ \hline
32\%    & Web and API serving                                                      \\ \hline
21\%    & Data Processing, e.g., batch ETL (database Extract, Transform, and Load) \\ \hline
17\%    & Integrating 3rd Party Services                                           \\ \hline
16\%    & Internal tooling                                                         \\ \hline
8\%     & Chat bots e.g., Alexa Skills (SDK for Alexa AI Assistant)                \\ \hline
6\%     & Internet of Things                                                       \\ \hline
\end{tabular}
}
\caption{Popularity of serverless computing use cases according to a 2018 survey\cite{serverless_growth}.}
\label{tbl:serverless_popularity}
\end{table}
\section{Limitations of Today's Serverless Computing Platforms}
\label{sec:sec3}

Serverless cloud functions have been successfully employed for several classes of workloads\footnote{See ``Use Cases'' here: \url{https://aws.amazon.com/lambda/}.} including  API serving, event stream processing, and limited ETL\footnote{The ETL implemented with today's cloud functions is typically restricted to Map-only processing.
} (see Table~\ref{tbl:services}). To see what obstacles prevent supporting more general workloads, we attempted to create serverless versions of applications that were of interest to us and studied examples published by others. These are not intended to be representative of the rest of information technology outside of the current serverless computing ecosystem; they are simply examples selected to uncover common weaknesses that might prevent serverless versions of many other interesting applications.

In this section, we present an overview of five research projects and discuss the obstacles that prevent existing serverless computing platforms from achieving state-of-the-art performance, i.e., matching the performance of serverful clouds for the same workloads. We are focused in particular on approaches that utilize general purpose cloud functions for compute, rather than relying heavily on other application-specific serverless offerings (BaaS). However in our final example, Serverless SQLite, we identify a use case that maps so poorly to FaaS that we conclude that databases and other state-heavy applications will remain as BaaS. An appendix at the end of this paper goes into more detail of each application. 

\renewcommand{\arraystretch}{1.5} 

\begin{table}[t!]
\centering
\resizebox{\columnwidth}{!}{%
\begin{tabular}{|>{\raggedright}p{2.1cm}|>{\raggedright}p{2.2cm}|>{\raggedright}p{3.2cm}|>{\raggedright}p{3cm}|>{\raggedright}p{3cm}|}
\hline
\textit{Application}                            & \textit{Description}                           & \textit{Challenges}                                                                                              & \textit{Workarounds}                                                                                       & \textit{Cost-performance}                                                                                                 \tabularnewline \hline
Real-time video compression (ExCamera) & On-the-fly video encoding             & Object store too slow to support fine-grained communication; functions too coarse grained for tasks.    & Function-to-function communication to avoid object store; a function executes more than one task. & 60x faster, 6x cheaper versus VM instances.                                                                       \tabularnewline \hline
MapReduce                              & Big data processing (Sort 100TB)      & Shuffle doesn't scale due to object store’s latency and IOPS limits                                     & Small storage with low-latency, high IOPS to speed-up shuffle.                                     & Sorted 100 TB 1\% faster than VM instances, costs 15\% more.                                                      \tabularnewline \hline
Linear algebra (Numpywren)             & Large scale linear algebra            & Need large problem size to overcome storage (S3) latency, hard to implement efficient broadcast.        & Storage with low-latency high-throughput to handle smaller problem sizes.                          & Up to 3x slower completion time. 1.26x to 2.5x lower in CPU resource consumption. \tabularnewline \hline
ML pipelines (Cirrus)                  & ML training at scale                  & Lack of fast storage to implement parameter server; hard to implement efficient broadcast, aggregation. & Storage with low-latency, high IOPS to implement parameter server.                                 & 
3x-5x faster than VM instances, up to 7x higher total cost. \tabularnewline \hline
Databases (Serverless SQLite)          & Primary state for applications (OLTP) & Lack of shared memory, object store has high latency, lack of support for inbound connectivity.         & Shared file system can work if write needs are low.                                                & 3x higher cost per transaction than published TPC-C benchmarks. Reads scale to match but writes do not.           \tabularnewline \hline
\end{tabular}
}
\caption{Summary of requirements for new application areas for serverless computing.}
\label{tbl:applications}
\end{table}

Interestingly, even this eclectic mix of applications exposed similar weaknesses, which we list after describing the applications. Table~\ref{tbl:applications} summarizes the five applications.

\textbf{ExCamera: Video encoding in real-time.} ExCamera~\cite{fouladi2017encoding} aims to provide a real-time encoding service to users uploading their videos to sites, such as YouTube. Depending on the size of the video, today's encoding solutions can take tens of minutes, even hours. To perform encoding in real time, ExCamera parallelizes the ``slow'' parts of the encoding, and performs the ``fast'' parts serially. ExCamera exposes the internal state of the video encoder and decoder, allowing encoding and decoding tasks to be executed using purely functional semantics. In particular, each task takes the internal state along with video frames as input, and emits the modified internal state as output.

\textbf{MapReduce.} Analytics frameworks such as MapReduce, Hadoop, and Spark, have been traditionally deployed on managed clusters. While some of these analytics workloads are now moving to serverless computing, these workloads mostly consist of Map-only jobs. The natural next step is supporting full fledged MapReduce jobs. One of the driving forces behind this effort is leveraging the flexibility of serverless computing to efficiently support jobs whose resource requirements vary significantly during their execution.

\textbf{Numpywren: Linear algebra.} Large scale linear algebra computations are traditionally deployed on supercomputers or high-performance computing clusters connected by high-speed, low-latency networks. Given this history, serverless computing initially seems a poor fit. Yet there are two reasons why serverless computing might still make sense for linear algebra computations. First, managing clusters is a big barrier for many non-CS scientists~\cite{jonas2017occupy}. Second, the amount of parallelism can vary dramatically during a computation. Provisioning a cluster with a fixed size will either slow down the job or leave the cluster underutilized. 

\textbf{Cirrus: Machine learning training.} Machine Learning (ML) researchers have traditionally used clusters of VMs for different tasks in ML workflows such as preprocessing, model training, and hyperparameter tuning. One challenge with this approach is that different stages of this pipeline can require significantly different amounts of resources. As with linear algebra algorithms, a fixed cluster size will either lead to severe underutilization or severe slowdown. Serverless computing can address this challenge by enabling every stage to scale to meet its resource demands. Further, it frees developers from managing clusters.

\textbf{Serverless SQLite:  Databases.} Various autoscaling database services already exist \cite{dynamodb, clouddatastore, aurora, cloudspanner, cosmosdb, faunadb}, but to better understand the limitations of serverless computing it is important to understand what makes database workloads particularly challenging to implement. In this context, we consider whether a third party could implement a serverless database directly using cloud functions, the general purpose serverless computing building block. A strawman solution would be to run common transactional databases, such as PostgreSQL, Oracle, or MySQL inside cloud functions. However, that immediately runs into a number of challenges. First, serverless computing has no built-in persistent storage, so we need to leverage some remote persistent store, which introduces large latency. Second, these databases assume connection-oriented protocols, e.g., databases are running as servers accepting connections from clients. This assumption conflicts with existing  cloud functions that are running behind network address translators, and thus don't support incoming connections. Finally, while many high performance databases rely on shared memory~\cite{hellerstein2007architecture}, cloud functions run in isolation so cannot share memory. While shared-nothing distributed databases~\cite{corbett2013spanner, cockroachdb, NuoDB} do not require shared memory, they expect nodes to remain online and be directly addressable. All these issues pose significant challenges to running traditional database software atop of serverless computing, or to implementing equivalent functionality, so we expect databases to remain BaaS.

One of the key reasons these applications hope to use serverless computing is fine-grained autoscaling, so that resource utilization closely matches each application's the varying demand. Table~\ref{tbl:applications} summarizes the characteristics, challenges, and workarounds for these five applications, which we next use to identify four limits in the current state of serverless computing.

\subsection{Inadequate storage for fine-grained operations}

The stateless nature of serverless platforms makes it difficult to support applications that have fine-grained state sharing needs. This is mostly due to the limitations of existing storage services offered by cloud providers. Table~\ref{tbl:storage} summarizes the properties of the existing cloud storage services.

Object storage services such as AWS S3, Azure Blob Storage, and Google Cloud Storage are highly scalable and provide inexpensive long-term object storage, but exhibit high access costs and high access latencies. According to recent tests, all these services take at least 10 milliseconds to read or write small objects~\cite{performance_unequal}. With respect to IOPS, after the recent limit increase~\cite{s3_increase_perf}, S3 provides high throughput, but it comes with a high cost. Sustaining 100K IOPS costs \$30/min~\cite{s3_pricing}, 3 to 4 orders of magnitude more than running an AWS ElastiCache instance~\cite{elasticcache_pricing}. Such an ElastiCache instance provides better performance along several axes, with sub-millisecond read and write latencies, and over 100K IOPS for one instance configured to run the single-threaded Redis server.

Key-value databases, such as AWS DynamoDB, Google Cloud Datastore, or Azure Cosmos DB provide high IOPS, but are expensive and can take a long time to scale up.\footnote{Official best practices for scaling Google Cloud Datastore include the ``500/50/5'' rule: start with 500 operations per second, then increase by 50\% every 5 minutes. \url{https://cloud.google.com/datastore/docs/best-practices}.} Finally, while cloud providers offer in-memory storage instances based on popular open source projects such as Memcached or Redis, they are not fault tolerant and do not autoscale as do serverless computing platforms. 

As can be seen in Table~\ref{tbl:applications}, applications built on serverless infrastructure require storage services with transparent provisioning, the storage equivalent of compute autoscaling. Different applications will likely motivate different guarantees of persistence and availability, and perhaps also latency or other performance measures. We believe this calls for the development of ephemeral and durable serverless storage options, which we discuss further in Section~\ref{sec:sec4}.

\subsection{Lack of fine-grained coordination}

To expand support to stateful applications, serverless frameworks need to provide a way for tasks to coordinate. For instance, if task $A$ uses task $B$'s output there must be a way for $A$ to know when its input is available, even if $A$ and $B$ reside on different nodes. Many protocols aiming to ensure data consistency also require similar coordination. 

None of the existing cloud storage services come with notification capabilities. While cloud providers do offer stand-alone notification services, such as SNS~\cite{pub_sub} and SQS~\cite{garfinkeltechnical}, these services add significant latency, sometimes hundreds of milliseconds. Also, they can be costly when used for fine grained coordination. There have been some proposed research systems such as Pocket~\cite{klimovic2018pocket} that do not have many of these drawbacks, but they have not yet been adopted by cloud providers.

As such, applications are left with no choice but to either (1) manage a VM-based system that provides notifications, as in ElastiCache and SAND~\cite{akkus2018sand}, or (2) implement their own notification mechanism, such as in ExCamera~\cite{fouladi2017encoding}, that enables cloud functions to communicate with each other via a long-running VM-based rendezvous server. This limitation also suggests that new variants of serverless computing may be worth exploring, for example naming function instances and allowing direct addressability for access to their internal state (e.g., Actors as a Service~\cite{elastic_database}). 

\newcolumntype{D}{ >{\centering\arraybackslash} m{2cm} }
\newcolumntype{T}{ >{\raggedright\arraybackslash} m{0.2cm} }
\newcolumntype{Z}{ >{\raggedright\arraybackslash} m{1cm} }

\newcommand{\specialcell}[2][c]{%
  \begin{tabular}[#1]{@{}l@{}}#2\end{tabular}}

\definecolor{lightred}{RGB}{224,102,102}
\definecolor{lightgreen}{rgb}{0.57, 0.76, 0.49}
\definecolor{lightorange}{RGB}{246,178,107}

\newcommand{\floatfootnote}[1]{\ifx\[$\else\tablefootnote{#1}\fi}

\renewcommand{\arraystretch}{1.5} 
\begin{table}[t!]
\centering
\resizebox{\columnwidth}{!}{%
\begin{tabular}{|T|Z|D|D|D|D|D|D|} 
\hline
\multicolumn{2}{|l|}{} & Block Storage (e.g., AWS EBS, IBM Block Storage)                    
& Object Storage (e.g., AWS S3, Azure Blob Store, Google Cloud Storage)
& File System (e.g., AWS EFS, Google Filestore)                 
& Elastic Database (e.g., Google Cloud Datastore, Azure Cosmos DB) & 
Memory Store (e.g., AWS ElastiCache, Google Cloud Memorystore)
& \textbf{``Ideal'' storage service for serverless \mbox{computing}}

\\ 
\hline
\multicolumn{2}{|l|}{\specialcell{Cloud functions access}}            & {\cellcolor{lightred}}No             & {\cellcolor{lightgreen}}Yes & {\cellcolor{lightgreen}}Yes\tablefootnote{A shared file system is integrated on Azure Functions, but not on other cloud functions platforms.} & {\cellcolor{lightgreen}}Yes & {\cellcolor{lightgreen}}Yes & {\cellcolor{lightgreen}}Yes\\ 
\hline
\multicolumn{2}{|l|}{\makecell[l]{Transparent\\Provisioning}}     & {\cellcolor{lightred}}No             & {\cellcolor{lightgreen}}Yes & {\cellcolor{lightorange}}Capacity only\tablefootnote{File system capacity scales automatically on Azure, AWS, IBM, but not on Google. IOPS do not scale independently of storage space for any provider's file system.} & {\cellcolor{lightgreen}}Yes\tablefootnote{Google Cloud Datastore and DynamoDB provisioned capacity autoscaling can be slow to respond to load spikes.} & {\cellcolor{lightred}}No & {\cellcolor{lightgreen}}Yes             \\ 
\hline
\multicolumn{2}{|l|}{\makecell[l]{Availability and\\persistence guarantees}}                  & {\cellcolor{lightorange}}Local \& Persistent  & {\cellcolor{lightgreen}}Distributed \& Persistent & {\cellcolor{lightgreen}}Distributed \& Persistent& {\cellcolor{lightgreen}}Distributed \& Persistent& {\cellcolor{lightred}}Local \& Ephemeral & {\cellcolor{lightorange}}Various            \\ 
\hline
\multicolumn{2}{|l|}{Latency (mean)}               & {\cellcolor{lightgreen}}$<1$ms  & {\cellcolor{lightred}}$10-20$ms            & {\cellcolor{lightorange}}$4-10$ms & {\cellcolor{lightred}}$8-15$ms & {\cellcolor{lightgreen}}$<1$ms & {\cellcolor{lightgreen}}$<1$ms \\ 
\hline
\parbox[t]{2mm}{\multirow{5}{*}{\rotatebox[origin=c]{90}{Cost\tablefootnote{For services that do not charge per operation, such as ElastiCache, we linearly scale down the pricing. For example, to calculate the cost of 1 IOPS, we take the cost of an instance that can sustain 30K IOPS with 4KB blocks and divide by 30K.}}}} &  \specialcell{Storage capacity\\ (1 GB/month)} & {\cellcolor{lightgreen}}\$0.10  & {\cellcolor{lightgreen}}\$0.023 & {\cellcolor{lightorange}}\$0.30 & {\cellcolor{lightorange}}\$0.18--\$0.25 & {\cellcolor{lightred}}\$1.87 & {\cellcolor{lightgreen}}$\sim$\$0.10             \\ 
\hhline{|~|-|-|-|-|-|-|-|}
                                & \specialcell{Throughput (1\\MB/s for 1 month)}       & {\cellcolor{lightgreen}}\$0.03 & {\cellcolor{lightgreen}}\$0.0071 & {\cellcolor{lightred}}\$6.00            & {\cellcolor{lightred}}\$3.15-\$255.1            & {\cellcolor{lightorange}}\$0.96 & {\cellcolor{lightgreen}}$\sim$\$0.03 \\ 
\hhline{|~|-|-|-|-|-|-|-|}
                                & \specialcell{IOPS\\(1/s for 1 month)}             & {\cellcolor{lightgreen}}\$0.03  & {\cellcolor{lightred}}\$7.1            & {\cellcolor{lightorange}}\$0.23 & {\cellcolor{lightred}}\$1-\$3.15        & {\cellcolor{lightgreen}}\$0.037 & {\cellcolor{lightgreen}}$\sim$\$0.03  \\
\hline
\end{tabular}
}
\caption{Characteristics of storage services by cloud providers to the serverless ideal. Costs are monthly values for storing 1 GB (capacity), transferring 1 MB/s (throughput), and issuing 1 IOPS (or 2.4 million requests in 30 days). All values reflect a 50/50 read/write balance and a minimum transfer size of 4 KB. The color codings of entries are green for good, orange for medium, and red for poor. Persistence and availability guarantees describe how well the system tolerates failures: local provides reliable storage at one site, distributed ensures the ability to survive site failures, and ephemeral describes data that resides in memory and is subject to loss, e.g., in the event of software crashes. The serverless ideal would provide cost and performance comparable to block storage, while adding transparent provisioning and access for cloud functions.}
\label{tbl:storage}
\end{table}

\subsection{Poor performance for standard communication patterns}

\begin{figure}[!t]
  \centering
  \includegraphics[width=5in]{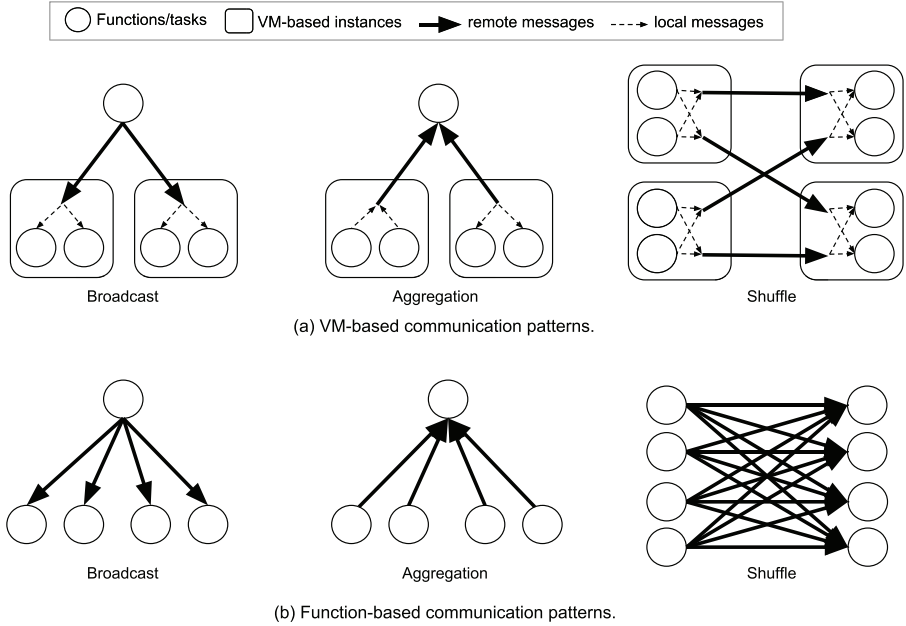}
  \caption{Three common communication patterns for distributed applications: broadcast, aggregation, and shuffle. (a) Shows these communication patterns for VM instances where each instance runs two functions/tasks. (b) Shows the same communication patterns for cloud function instances. Note the significantly lower number of remote messages for the VM-based solutions. This is because VM instances offer ample opportunities to share, aggregate, or combine data locally across tasks before sending it or after receiving it.}
  \label{fig:std_comm_patterns}
\end{figure}

Broadcast, aggregation, and shuffle are some of the most common communication primitives in distributed systems. These operations are commonly employed by applications such as machine learning training and big data analytics. Figure~\ref{fig:std_comm_patterns} shows the communication patterns for these primitives for both VM-based and function-based solutions. 

With VM-based solutions, all tasks running on the same instance can share the copy of the data being broadcast, or perform local aggregation before sending the partial result to other instances. As such, the communication complexity of the broadcast and aggregation operations is $O(N)$, where $N$ is the number of VM-instances in the system. However, with cloud functions this complexity is $O(N \times K)$, where $K$ is the number of functions per VM. The difference is even more dramatic for the shuffle operation. With VM-based solutions all local tasks can combine their data such that there is only one message between two VM instances. Assuming the same number of senders and receivers yields $N^2$ messages. For comparison, with cloud functions we need to send $(N \times K)^2$ messages. As existing functions have a much smaller number of cores than a VM, $K$ typically ranges from 10 to 100. Since the application cannot control the location of the cloud functions, a serverless computing application may need to send two and four orders of magnitude more data than an equivalent VM-based solution.

\subsection{Predictable Performance}

Although cloud functions have a much lower startup latency than traditional VM-based instances, the delays incurred when starting new instances can be high for some applications. There are three factors impacting this “cold start” latency: (1) the time it takes to start a cloud function; (2) the time it takes to initialize the software environment of the function, e.g., load Python libraries; and (3) application-specific initialization in user code. The latter two can dwarf the former. While it can take less than one second to start a cloud function, it might take tens of seconds to load all application libraries.\footnote{Consider a cloud function that needs to load a few hundred MB worth of Python libraries from object storage.}

Another obstacle to predictable performance is the variability in the hardware resources that results from giving the cloud provider flexibility to choose the underlying server. In our experiments~\cite{shankar2018numpywren}, we sometimes received CPUs from different hardware generations. This uncertainty exposes a fundamental tradeoff between the cloud provider desire to maximize the use of their resources and predictability.

\section{What Serverless Computing Should Become}
\label{sec:sec4}
Now that we've explained today's serverless computing and its limitations, let's look to the future to understand how to bring its advantages to more applications. Researchers have already begun to address issues raised above and to explore how to improve serverless platforms and the performance of workloads that run on them \cite{hendrickson2016serverless, baldini2017serverless}. Additional work done by our Berkeley colleagues and some of us emphasizes data-centric, distributed systems, machine learning, and programming model challenges and opportunities for serverless computing \cite{hellerstein2018serverless}. Here we take a broad view on increasing the types of applications and hardware that work well with serverless computing, identifying research challenges in five areas: abstractions, systems, networking, security, and architecture.

\subsection{Abstraction challenges}
\label{subsec:abstractionchallenges}
\textbf{Resource requirements}: With today's serverless offerings the developer specifies the cloud function's memory size and execution time limit, but not other resource needs. This abstraction hinders those who want more control on specifying resources, such as the number of CPUs, GPUs, or other types of accelerators. One approach would be to enable developers to specify these resource requirements explicitly. However, this would make it harder for cloud providers to achieve high utilization through statistical multiplexing, as it puts more constraints on function scheduling. It also goes against the spirit of serverless by increasing the management overhead for cloud application developers.

A better alternative would be to raise the level of abstraction, having the cloud provider infer resource requirements instead of having the developer specify them. To do so, the cloud provider could use a variety of approaches from static code analysis, to profiling previous runs, to dynamic (re)compilation to retarget the code to other architectures. Provisioning just the right amount of memory automatically is particularly appealing but especially challenging when the solution must interact with the automated garbage collection used by high-level language runtimes. Some research suggests that these language runtimes could be integrated with serverless platforms \cite{larisch2018alto}.

\textbf{Data dependencies}: Today's cloud function platforms have no knowledge of the data dependencies between the cloud functions, let alone the amount of data these functions might exchange. This ignorance can lead to suboptimal placement that could result in inefficient communication patterns, as illustrated in the MapReduce and numpywren examples (see Section~\ref{sec:sec3} and Appendix).

One approach to address this challenge would be for the cloud provider to expose an API that allows an application to specify its computation graph, enabling better placement decisions that minimize communication and improve performance. We note that many general purpose distributed frameworks (e.g., MapReduce, Apache Spark and Apache Beam/Cloud Dataflow \cite{akidau2015dataflow}), parallel SQL engines (e.g., BigQuery, Azure Cosmos DB), as well as orchestration frameworks (e.g., Apache Airflow \cite{airflow}) already produce such computation graphs internally. In principle, these systems could be modified to run on cloud functions and expose their computation graphs to the cloud provider. Note that AWS Step Functions represents progress in this direction by providing a state machine language and API.

\subsection{System challenges}
\textbf{High-performance, affordable, transparently provisioned storage}: As discussed in Section~\ref{sec:sec3} and Table~\ref{tbl:applications}, we see two distinct unaddressed storage needs: \textit{Serverless Ephemeral Storage} and \textit{Serverless Durable Storage}.

\textit{Ephemeral Storage}. The first four applications from Section~\ref{sec:sec3} were limited by the speed and latency of the storage system used to transfer state between cloud functions. While their capacity demands vary, all need such storage to maintain application state during the application lifetime. Once the application finishes, the state can be discarded. Such ephemeral storage might also be configured as a cache in other applications.

One approach to providing ephemeral storage for serverless applications would be to build a distributed in-memory service with an optimized network stack that ensures microsecond-level latency. This system would enable the functions of an application to efficiently store and exchange state during the application's lifetime. Such an in-memory service would need to automatically scale the storage capacity and the IOPS with the application's demands. A unique aspect of such a service is that it not only needs to \textit{allocate} memory transparently, but also \textit{free} it transparently. In particular, when the application terminates or fails, the storage allocated to that application should be automatically released. This management is akin to the OS automatically freeing the resources allocated by a process when the process finishes (or crashes). Furthermore, such storage must provide access protection and performance isolation across applications.

RAMCloud \cite{ousterhout2015ramcloud} and FaRM \cite{dragojevic2014farm} show that it is possible to build in-memory storage systems that can provide microsecond level latencies and support hundred of thousands IOPS per instance. They achieve this performance by carefully optimizing the entire software stack and by leveraging RDMA to minimize latency. However, they require applications to provision storage explicitly. They also do not provide strong isolation between multiple tenants. Another recent effort, Pocket \cite{klimovic2018pocket}, aims to provide the abstraction of ephemeral storage, but also lacks autoscaling, requiring applications to allocate storage a priori.

By leveraging statistical multiplexing, this ephemeral storage can be more memory-efficient than today's serverful computing. With serverful computing, if an application needs less memory than the aggregated memory of the allocated VM instances, that memory goes to waste. In contrast, with a shared in-memory service, any memory not used by one serverless application can be allocated to another. In fact, statistical multiplexing can benefit even a single application: with serverful computing, the unused memory of a VM cannot be used by the program running on another VM belonging to the same application, while in the case of a shared in-memory service it can. Of course, even with serverless computing there can be internal fragmentation if the cloud function doesn't use its entire local memory. In some cases, storing the application state of cloud functions in a shared in-memory service could alleviate the consequences of internal memory fragmentation. 

\textit{Durable Storage}. Like the other applications, our serverless  database application experiment was limited by the latency and IOPS of the storage system, but it also required long term data storage and the mutable-state semantics of a file system. While it's likely that database functionality, including OLTP, will increasingly be provided as a BaaS offering,\footnote{One recent example is Amazon Aurora Serverless (\url{https://aws.amazon.com/rds/aurora/serverless/}). Services such as Google's BigQuery and AWS Athena are basically serverless query engines rather than fully fledged databases.} we see this application as representative of several applications that require longer retention and greater durability than the Serverless Ephemeral Storage is suited to provide. To implement high-performance Serverless Durable Storage, one approach would be to leverage an SSD-based distributed store paired with a distributed in-memory cache. A recent system realizing many of these goals is the Anna key-value database that achieves both cost efficiency and high performance by combining multiple existing cloud storage offerings \cite{wu2018eliminating}. A key challenge with this design is achieving low tail latency in the presence of heavy tail access distributions, given the fact that in-memory cache capacity is likely to be much lower than SSD capacity. Leveraging new storage technologies \cite{chen2016review}, which promise microsecond-level access times, is emerging as a promising approach to address this challenge.

Similar to Serverless Ephemeral Storage, this service should be transparently provisioned and should ensure isolation across applications and tenants for security and predictable performance. However, whereas Serverless Ephemeral Storage would reclaim resources when an application terminates, Serverless Durable Storage must only free resources explicitly (e.g., as a result of a ``delete'' or ``remove'' command), just like in traditional storage systems. Additionally, it must of course ensure durability, so that any acknowledged writes will survive failures.  

\textbf{Coordination/signaling service}: Sharing state between functions often uses a producer-consumer design pattern, which requires consumers to know as soon as the data is available from producers. Similarly, one function might want to signal another when a condition becomes available, or multiple functions might want to coordinate, e.g., to implement data consistency mechanisms. Such signaling systems would benefit from microsecond-level latency, reliable delivery, and broadcast or group communication. We also note that since cloud function instances are not individually addressable they cannot be used to implement textbook distributed systems algorithms such as consensus or leader election \cite{hellerstein2018serverless}.

\textbf{Minimize startup time}: There are three parts of startup time (1) scheduling and starting resources to run the cloud function, (2) downloading the application software environment (e.g., operating system, libraries) to run the function code, and (3) performing application-specific startup tasks such as loading and initializing data structures and libraries. Resource scheduling and initialization can incur significant delays and overheads from creating an isolated execution environment, and from configuring customer's VPC and IAM policies. Cloud providers \cite{firecracker, gVisor}, as well as others \cite{binaris,oakes2018sock} have recently focused on reducing the startup time by developing new lightweight isolation mechanisms.

One approach to reduce (2) is leveraging unikernels \cite{manco2017my}. Unikernels obviate the overhead incurred by traditional operating systems in two ways. First, instead of \textit{dynamically} detecting the hardware, applying user configurations, and allocating data structures like traditional operating systems, unikernels squash these costs by being preconfigured for the hardware they are running on and statically allocating the data structures. Second, unikernels include only the drivers and system libraries strictly required by the application, which leads to a much lower footprint than traditional operating systems. It is worth noting that since unikernels are tailored to specific applications, they cannot realize some of the efficiencies possible when running many instances of a standardized kernel, for example sharing kernel code pages between different cloud functions on the same VM, or reducing the start-up time by pre-caching. Another approach to reduce (2) is to dynamically and incrementally load the libraries as they are invoked by the application, for example as enabled by the shared file system used in Azure Functions. 

Application-specific initialization (3) is the responsibility of the programmer, but cloud providers can include a readiness signal in their API to avoid sending work to function instances before they can start processing it \cite{knative}. More broadly, cloud providers can seek to perform startup tasks ahead of time \cite{oakes2018sock}. This is particularly powerful for customer-agnostic tasks such as booting a VM with popular operating system and set of libraries, as a ``warm pool'' of such instances can be shared between tenants \cite{wagnercompute}.

\subsection{Networking challenges}
As explained in Section~\ref{sec:sec3} and as illustrated in Figure~2, cloud functions can impose significant overhead on popular communication primitives such as broadcast, aggregation, and shuffle. In particular, assuming that we can pack $K$ cloud functions on a VM instance, a cloud function version would send $K$ times more messages than an instance version, and $K^2$ more messages in the case of shuffle.

There may be several ways to address this challenge:
\begin{itemize}[noitemsep]
\item Provide cloud functions with a larger number of cores, similar to VM instances, so multiple tasks can combine and share data among them before sending over the network or after receiving it. 
\item Allow the developer to explicitly place the cloud functions on the same VM instance. 
Offer distributed communication primitives that applications can use out-of-the-box so that cloud providers can allocate cloud functions to the same VM instance. 
\item Let applications provide a computation graph, enabling the cloud provider to co-locate the cloud functions to minimize communication overhead (see ``Abstraction Challenges'' above.)
\end{itemize}
Note that the first two proposals could reduce the flexibility of cloud providers to place cloud functions, and consequently reduce data center utilization. Arguably, they also go against the spirit of serverless computing, by forcing developers to think about system management.

\subsection{Security challenges}
Serverless computing reshuffles security responsibilities, shifting many of them from the cloud user to the cloud provider without fundamentally changing them. However, serverless computing must also grapple with the risks inherent in both application disaggregation multi-tenant resource sharing. 

\textbf{Scheduling randomization and physical isolation}: Physical co-residency is the center of hardware-level side-channel or Rowhammer~\cite{kim2014flipping} attacks inside the cloud. As a first step in these types of attacks, the adversarial tenant needs to confirm the cohabitation with the victim on the same physical host, instead of randomly attacking strangers. The ephemerality of cloud functions may limit the ability of the attacker to identify concurrently-running victims. A randomized, adversary-aware scheduling algorithm~\cite{ristenpart2009hey} might lower the risk of co-locating the attacker and the victim, making co-residency attacks more difficult. However, deliberately preventing physical co-residency may conflict with placement to optimize start-up time, resource utilization, or communication. 

\textbf{Fine-grained security contexts}: Cloud functions need fine-grained configuration, including access to private keys, storage objects, and even local temporary resources. There will be requirements for translating security policies from existing serverful applications, and for offering highly-expressive security APIs for dynamic use in cloud functions. For example, a cloud function may have to delegate security privileges to another cloud function or cloud service. A capability-based access control mechanism using cryptographically protected security contexts could be a natural fit for such a distributed security model. Recent work \cite{alpernas2018secure} suggests using information flow control for cross-function access control in a multi-party setting. Other challenges of providing distributed management of security primitives, such as non-equivocation \cite{eschenauer2002key} and revocation, are exacerbated if short-lived keys and certificates are dynamically created for cloud functions.

At the system level, users demand more fine-grained security isolation for each function, at least as an option. The challenge in providing function-level sandboxing is to maintain a short startup time without caching the execution environments in a way that shares state between repeated function invocations. One possibility would be to locally snapshot the instances so that each function can start from clean state. Alternatively, light-weight virtualization technologies are starting to be adopted by serverless providers: library OSes, including gVisor \cite{gVisor}, implement system APIs in a user-space ``shim layer,'' while unikernels and microVMs, including AWS Firecracker \cite{firecracker}, streamline the guest kernels and help minimize the host attack surface. These isolation techniques reduce startup times to as little as tens of milliseconds, as compared to VM startup times measured in seconds. Whether these solutions achieve parity to traditional VMs in terms of security remains to be shown, and we expect the search for strong isolation mechanisms with low startup overheads to be an active area of ongoing research and development. On the positive side, provider management and short-lived instances in serverless computing can enable much faster patching of vulnerabilities.

For users who want protection against co-residency attacks, one solution would be demanding physical isolation. Recent hardware attacks (e.g., Spectre~\cite{Kocher2018spectre} and Meltdown~\cite{Lipp2018meltdown}) also make reserving a whole core or even a whole physical machine appealing for users.  Cloud providers may offer a premium option for customers to launch functions on physical hosts dedicated exclusively to their use.

\textbf{Oblivious serverless computing}: Cloud functions can leak access patterns and timing information through communication. For serverful applications, data is usually retrieved in a batch, and cached locally. In contrast, because cloud functions are ephemeral and widely distributed across the cloud, the network transmission patterns can leak more sensitive information to a network attacker in the cloud (e.g., an employee), even if the payload is encrypted end-to-end. The tendency to decompose serverless applications into many small functions exacerbates this security exposure. While the primary security concern is from external attackers, the network patterns can be protected from employees by adopting oblivious algorithms. Unfortunately, these tend to have high overhead \cite{shi2011oblivious}. 

\subsection{Computer architecture challenges}
\textbf{Hardware Heterogeneity, Pricing, and Ease of Management}: Alas, the x86 microprocessors that dominate the cloud are barely improving in performance. In 2017, single program performance improvement only 3\% \cite{hennessy2011computer}. Assuming the trends continue, performance won't double for 20 years. Similarly, DRAM capacity per chip is approaching its limits; 16 Gbit DRAMs are for sale today, but it appears infeasible to build a 32 Gbit DRAM chip. A silver lining of this slow rate of change is letting providers replace older computers as they wear out with little disruption to the current serverless marketplace.

Performance problems for general purpose microprocessors do not reduce the demand for faster computation. There are two paths forward \cite{hennessy2018new}. For functions written in high-level scripting languages like JavaScript or Python, hardware-software co-design could lead to language-specific custom processors that run one to three orders of magnitude faster. The other path forward is \textit{Domain Specific Architectures}. DSAs are tailored to a specific problem domain and offer significant performance and efficiency gains for that domain, but perform poorly for applications outside that domain. Graphical Processing Units (GPUs) have long been used to accelerate graphics, and we're starting to see DSAs for machine learning such as the Tensor Processing Units (TPUs). TPUs can outperform CPUs by a factor of 30x. These examples are the first of many, as general purpose processors enhanced with DSAs for separate domains will become the norm.

As mentioned above in Section~\ref{subsec:abstractionchallenges}, we see two paths for serverless computing to support the upcoming hardware heterogeneity: 
\begin{enumerate}[noitemsep]
\item Serverless could embrace multiple instance types, with a different price per accounting unit depending on the hardware used. 
\item The cloud provider could select language-based accelerators and DSAs automatically. This automation might be done implicitly based on the software libraries or languages used in a cloud function, say GPU hardware for CUDA code and TPU hardware for TensorFlow code. Alternatively, the cloud provider could monitor the performance of the cloud functions and migrate them to the most appropriate hardware the next time they are run. 
\end{enumerate}

Serverless computing is facing heterogeneity now in a small way for the SIMD instructions of the x86. AMD and Intel rapidly evolve that portion of the x86 instruction set by increasing the number of operations performed per clock cycle and by adding new instructions. For programs that use SIMD instructions, running on a recent Intel Skylake microprocessor with 512-bit wide SIMD instructions can be much faster than running on the older Intel Broadwell microprocessor with 128-bit wide SIMD instructions. Today both microprocessors are supplied at the same price in AWS Lambda, but there is currently no way for serverless computing users to indicate that they want the faster SIMD hardware. It seems to us that compilers should suggest which hardware would be the best match.

As accelerators become more popular in the cloud,  serverless cloud providers will no longer be able to ignore the dilemma of heterogeneity, especially since plausible remedies exist.

\section{Fallacies and Pitfalls}
\label{sec:sec5}

This section uses the fallacy and pitfall style of Hennessy and Patterson \cite{hennessy2011computer}.
\begin{description}[style=unboxed,leftmargin=0cm]
\item[\textbf{Fallacy}] \textit{Since an AWS Lambda cloud function instance with equivalent memory capacity of an on-demand AWS t3.nano instance (0.5 GiB) costs 7.5x as much per minute, serverless cloud computing is more expensive than serverful cloud computing.}
\end{description}
The beauty of serverless computing is all the system administration capability that is included in the price, including redundancy for availability, monitoring, logging, and scaling. Cloud providers report that customers see cost savings of 4x-10x when moving applications to serverless \cite{wagnerserverless}. The equivalent functionality is much more than a single t3.nano instance, which along with being a single point of failure operates with credit system that limits it to at most 6 minutes of CPU use per hour (5\% of the two vCPUs), so it could deny service during a load spike that serverless would easily handle. Serverless is accounted for at much finer boundaries, including for scaling up and down, and so may be more efficient in the amount of computing used. Because there is no charge when there are no events that invoke cloud functions, it's possible that serverless could be much less expensive.
\begin{description}[style=unboxed,leftmargin=0cm]
\item[\textbf{Pitfall}] \textit{Serverless computing can have unpredictable costs.}
\end{description}
For some users, a disadvantage of the pure pay-as-you-go model employed by serverless computing is cost unpredictability, which is at odds with the way many organizations manage their budgets. When approving the budget, which typically occurs annually, organizations want to know how much serverless services will cost over the next year. This desire is a legitimate concern, one which cloud providers might mitigate by offering bucket-based pricing, similar to the way phone companies offer fixed rate plans for certain amounts of usage. We also believe that as organizations use serverless more and more, they will be able to predict their serverless computing costs based on history, similar to the way they do today for other utility services, such as electricity.
\begin{description}[style=unboxed,leftmargin=0cm]
\item[\textbf{Fallacy}] \textit{Since serverless computing programming is in high-level languages like Python, it's easy to port applications between serverless computing providers.}
\end{description}
Not only do function invocation semantics and packaging differ between cloud serverless computing providers, but many serverless applications also rely upon an ecosystem of proprietary BaaS offerings that lacks standardization. Object storage, key-value databases, authentication, logging, and monitoring are prominent examples. To achieve portability, serverless users will have to propose and embrace some kind of standard API, such POSIX tried to do for operating systems. The Knative project from Google is a step in this direction, aiming to provide a unified set of primitives for application developers to use across deployment environments \cite{knative}.
\begin{description}[style=unboxed,leftmargin=0cm]
\item[\textbf{Pitfall}] \textit{Vendor lock-in may be stronger with serverless computing than for serverful computing.}
\end{description}
This pitfall is a consequence of the previous fallacy; if porting is hard, then vendor lock-in is likely.  Some frameworks promise to mitigate such lock-in with cross-cloud support \cite{serverlessinc}.
\begin{description}[style=unboxed,leftmargin=0cm]
\item[\textbf{Fallacy}] \textit{Cloud functions cannot handle very low latency applications needing predictable performance.}
\end{description}
The reason serverful instances handle such low-latency applications well is because they are always on, so they can quickly reply to requests when they receive them.  We note that if the start-up latency of a cloud function is not good enough for a given application, one could use a similar strategy: pre-warm cloud functions by exercising them regularly to ensure that there are enough running at any given time to satisfy the incoming requests.
\begin{description}[style=unboxed,leftmargin=0cm]
\item[\textbf{Pitfall}] \textit{Few so called ``elastic'' services match the real flexibility demands of serverless computing.}
\end{description}
The word ``elastic'' is a popular term today, but it is being applied to services that do not scale nearly as well as the best serverless computing services. We are interested in services which can change their capacity \textit{rapidly}, with \textit{minimal user intervention}, and can potentially ``scale to zero'' when not in use. For example, despite its name, AWS ElastiCache only allows you to instantiate an integral number of Redis instances. Other ``elastic''  services require explicit capacity provisioning, with some taking many minutes to respond to changes in demand, or scaling over only a limited range. Users lose many of the benefits of serverless computing when they build applications that combine highly-elastic cloud functions with databases, search indexes, or serverful application tiers that have only limited elasticity. Without a quantitative and broadly accepted technical definition or metric\textemdash{}something that could aid in comparing or composing systems\textemdash{}``elastic'' will remain an ambiguous descriptor.

\section{Summary and Predictions}
\label{sec:sec6}

By providing a simplified programming environment, serverless computing makes the cloud much easier to use, thereby attracting more people who can and will use it. Serverless computing comprises FaaS and BaaS offerings, and marks an important maturation of cloud programming. It obviates the need for manual resource management and optimization that today's serverful computing imposes on application developers, a maturation  akin to the move from assembly language to high-level languages more than four decades ago. 

We predict that serverless use will skyrocket. We also project that hybrid cloud on-premises applications will dwindle over time, though some deployments might persist due to regulatory constraints and data governance rules.

While already a success, we identified a few challenges that if overcome will make serverless popular for an even broader range of applications. The first step is Serverless Ephemeral Storage, which must provide low latency and high IOPS at reasonable cost, but need not provide economical long term storage. A second class of applications would benefit from Serverless Durable Storage, which does demand long term storage. New non-volatile memory technologies may help with such storage systems. Other applications would benefit from a low latency signaling service and support for popular communication primitives.

Two challenges for the future of serverless computing are improved security and accommodating cost-performance advances that are likely to come from special purpose processors.  In both cases, serverless computing has features that may help in addressing these challenges. Physical co-residency is a requirement for side-channel attacks, but it is much harder to confirm in serverless computing, and steps could be easily taken to randomize cloud function placement. The programming of cloud functions in high-level languages like JavaScript,  Python, or TensorFlow \cite{abadi2016tensorflow} raises the level of programming abstraction and makes it easier to innovate so that the underlying hardware can deliver improved cost-performance.

The Berkeley View of Cloud Computing paper~\cite{Armbrust09abovethe} projected that the challenges facing the cloud in 2009 would be addressed and that it would flourish, which it has. The cloud business is growing 50\% annually and is proving highly profitable for cloud providers.\footnote{Amazon Web Services (AWS) and Microsoft Azure are the largest cloud providers. According to a recent report in 2018 AWS had 41.5\% of ``application workloads'' in the public cloud, while Azure had 29.4\% (\url{https://www.skyhighnetworks.com/cloud-security-blog/microsoft-azure-closes-iaas-adoption-gap-with-amazon-aws/}). In term of revenue, Azure had a  \$37.6 billion run rate while AWS had a \$29.5 billion run rate (\url{https://techcrunch.com/2019/02/01/aws-and-microsoft-reap-most-of-the-benefits-of-expanding-cloud-market/}).}

We conclude this paper with the following predictions about serverless computing in the next decade:
\begin{itemize}[noitemsep]
\item We expect new BaaS storage services to be created that expand the types of applications that run well on serverless computing. Such storage will match the performance of local block storage and come in ephemeral and durable variants.
We will see much more heterogeneity of computer hardware for serverless computing than the conventional x86 microprocessor that powers it today.
\item We expect serverless computing to become simpler to program securely than serverful computing, benefiting from the high level of programming abstraction and the fine-grained isolation of cloud functions.
\item We see no fundamental reason why the cost of serverless computing should be higher than that of serverful computing, so we predict that billing models will evolve so that almost any application, running at almost any scale, will cost no more and perhaps much less with serverless computing. 
\item The future of serverful computing will be to facilitate BaaS.  Applications that prove to be difficult to write on top of serverless computing, such as OLTP databases or communication primitives such as queues, will likely be offered as part of a richer set of services from all cloud providers.
\item While serverful cloud computing won't disappear, the relative importance of that portion of the cloud will decline as serverless computing overcomes its current limitations.
\item Serverless computing will become the default computing paradigm of the Cloud Era, largely replacing serverful computing and thereby bringing closure to the Client-Server Era.  
\end{itemize}

\section{Acknowledgements}
\label{sec:sec7}

We'd like to thank people who gave feedback on early drafts of this paper: Michael Abd-El-Malek (Google), Aditya Akella (Wisconsin), Remzi H. Arpaci-Dusseau (Wisconsin), Bill Bolosky (Microsoft), Forrest Brazeal (Trek10), Eric Brewer (Google/UC Berkeley), Armando Fox (UC Berkeley), Changhua He (Ant Financial), Joseph M. Hellerstein (UC Berkeley), Mike Helmick (Google), Marvin Theimer (Amazon Web Services), Keith Winstein (Stanford), and Matei Zaharia (Stanford). The sponsors of the RISELab are Alibaba Group, Amazon Web Services, Ant Financial, Arm Holdings, Capital One, Ericsson, Facebook, Google, Huawei, Intel, Microsoft, Scotiabank, Splunk, VMware, and the National Science Foundation.
\bibliographystyle{unsrt}
\bibliography{bibtex}
\pagebreak{}
\section{Appendix. More Depth on Five Applications that Stretch Today's Serverless Computing}
\label{sec:sec8}

\subsection{ExCamera: Video encoding in real-time}
ExCamera \cite{fouladi2017encoding} aims to provide a real-time encoding service to users uploading their videos to sites, such as YouTube. Depending on the size of the video, today's encoding solutions can take tens of minutes, even hours. To perform encoding in real time, ExCamera parallelizes the ``slow'' parts of the encoding, and performs the ``fast'' parts serially. To do so, ExCamera exposes the internal state of the video encoder and decoder, allowing encoding and decoding tasks to be executed using purely functional semantics. In particular, each task takes the internal state along with video frames as input, and emits the modified internal state as output. 

ExCamera leverages the AWS Lambda platform to exploit of the new algorithm's parallelism so as to quickly scale up the computation. Unlike VM-based instances that take minutes to start and need to be explicitly managed, cloud functions start in seconds and require no management. 

Unfortunately, just using cloud functions out of the box is insufficient. The ExCamera tasks are fine-grained, taking as little as a few seconds, and operate on a non-trivial amount of state. Thus, using S3 to exchange the intermediate state incurs a significant overhead. Moreover, cloud functions are behind network address translators (NATs); they can initiate connections, but not accept them. This makes direct communication between cloud functions challenging. To get around this issue, ExCamera uses a rendezvous server to relay packets between cloud functions. Furthermore, ExCamera employs a coordinator to orchestrate the tasks and the communication between tasks across cloud functions. This support enables one function invocation to run multiple tasks, thereby amortizing its startup time. By doing so, ExCamera offers much lower latency than existing encoding (60x faster than Google's multithreaded vpxenc encoder on 128 cores) while still remaining cost effective (6.17x cheaper than encoding using vpxenc on an 128-core x1.32xlarge EC2 instance).

\subsection{MapReduce}
Analytics frameworks such as MapReduce, Hadoop, and Spark, have been traditionally deployed on managed clusters. While some of these analytics workloads are now moving to serverless computing, these workloads mostly consist of Map-only jobs. The natural next step is supporting full fledged MapReduce jobs. One of the driving forces behind this effort is leveraging the flexibility of serverless computing to efficiently support jobs whose resource requirements vary significantly during their execution. For example, query 95 in the TPC-DS benchmark \cite{tpcdsbenchmark} consists of eight stages, which process from 0.8 MB to 66 GB of data per stage, an almost five order of magnitude difference! 

\begin{figure}[!t]
  \centering
  \includegraphics[width=\columnwidth]{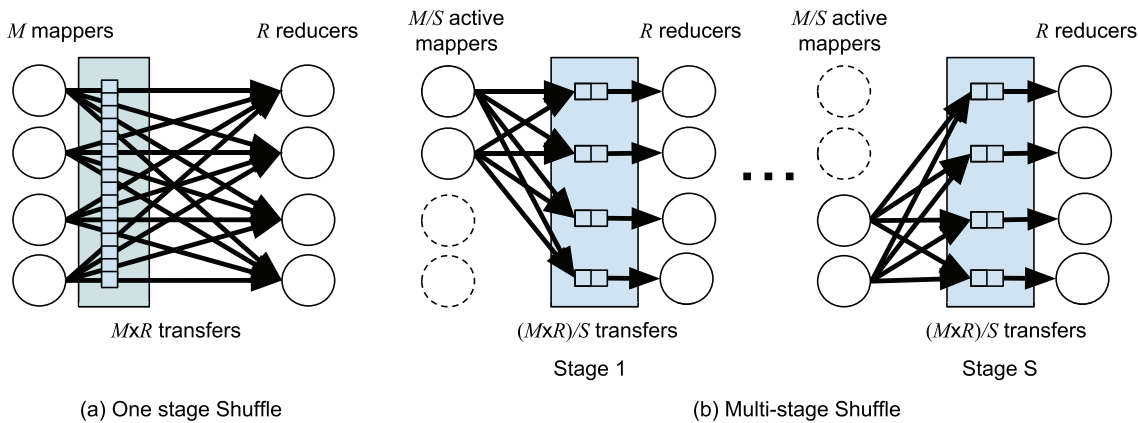}
  \caption{(a) Shuffle operation with M mappers and N receivers. Each transfer happens via external storage, which is shown by a large blue rectangle. The data corresponding to each transfer is shown by small squares.  (b) Multi-stage shuffle with S stages.}
  \label{fig:serverless_comm_patterns}
\end{figure}

The main challenge in supporting MapReduce jobs on top of the existing cloud functions platforms is the shuffle operation. With shuffle, every Map task sends data to every reduce task. Assuming there are $M$ mappers and $R$ reducers, a shuffle will generate $M \times R$ transfers (see Figure~\ref{fig:serverless_comm_patterns}(a)). Since cloud functions don't communicate directly with each other, all these transfers must take place via an external storage. The number of transfers can be large. 

As an example, assume we need to shuffle 100 TB of data with AWS cloud functions using S3 as the external storage. Given the existing resource constraints, a cloud function can process data blocks no larger than 3 GB, which is the largest memory capacity of a cloud function today  (see Table~\ref{tbl:serverless_vs_serverful}).\footnote{This assumes the cloud function needs to read all data before creating the output.} Hence, we need to partition the input into 33,000 blocks. If we were to have one Map function per block, and an equal number of reduce function (i.e., $M = R = 33,000$), we would need to perform 1.11 billion transfers, or 2.22 billion IO operations (i.e., one write and one read per transfer)! This number is significant for systems like S3 that limit the number of IO operations/sec (IOPS) and charge a premium per IO request. Thus, shuffling 100 TB can take tens of hours and cost \$12,000 for S3 IOPS alone. 

One alternative is using high performance storage, such as ElastiCache\footnote{Elasticache is based on Redis. A single single-threaded Redis instance can handle 100K+ IOPS. 100 instances can handle 1.3B transfers in just 130 seconds.} instead of S3. However, this high performance storage is expensive. Using 100 TB of such storage to shuffle 100 TB of data would be far more expensive than a VM-based solution. Fortunately, dividing the shuffle in stages (as in Figure~\ref{fig:serverless_comm_patterns}(b)) reduces the needed amount of high performance storage significantly. For example, if we use $S = 50$ stages for a 100 TB shuffle, we need only 2 TB of high-performance storage. By appropriately choosing the size of this storage, we come close to matching the existing VM-based frameworks in performance and cost. 

For example, the current record of 100 TB CloudSort benchmark was 2,983 seconds for \$144 using a cluster of 395 VMs, each with 4vCPU cores and 8GB memory. Our solution runs the same task in 2,945 seconds for \$163 using cloud functions (\$117 with AWS Lambda) and a hybrid of cloud object storage (\$14 for AWS S3 IOPS cost) and the ElastiCache service (\$32 for Redis cost).

\subsection{Numpywren: Linear algebra}
Large scale linear algebra computations are traditionally deployed on supercomputers or high-performance computing clusters connected by high-speed, low-latency networks. Given this history, serverless computing initially seems a poor fit. 

\begin{figure}[!t]
  \centering
  \includegraphics[scale=0.5]{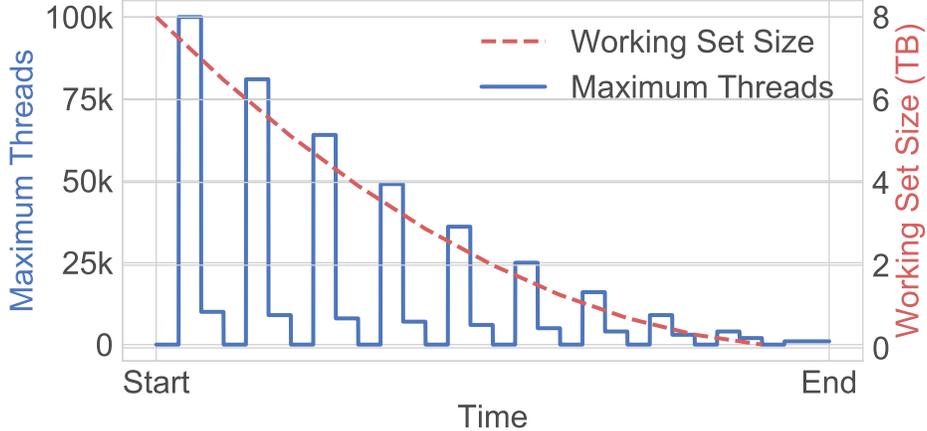}
  \caption{Theoretical profile of task parallelism and working set size over time in a distributed Cholesky decomposition.}
  \label{fig:theoretical_flops}
\end{figure}

Yet there are two reasons why serverless computing might still make sense for linear algebra computations. First, managing clusters is a big barrier for many non-CS scientists. Second, the amount of parallelism can vary dramatically during a computation. Figure~\ref{fig:theoretical_flops} shows the working set and the maximum degree of parallelism for Cholesky decomposition, one of the most popular methods for solving linear equation on a large matrix. Provisioning a cluster with a fixed size will either slow down the job or leave the cluster underutilized. 

Recent work on the numpywren project shows that serverless computing may be a good match for large scale linear algebra \cite{shankar2018numpywren}. The key insight is that, for many linear algebra operations, computation time often dominates communication for large problem sizes. For example, Cholesky, LU, and Singular Value decompositions, all exhibit $O(n^3)$ computation and $O(n^2)$ communication complexities. Numpywren leverages this observation, and shows for many popular algorithms that even when using high latency storage systems such as S3, a serverless computing implementation can achieve comparable performance to an optimized MPI implementation (ScaLAPACK) running on a dedicated cluster. For certain linear algebra algorithms such as matrix multiply and Cholesky decomposition and singular value decomposition, numpywren's performance (completion time) is only 1.3x (for all algorithms) of ScaLAPACK, and its CPU consumption (total CPU-hours) is 1.3x for matrix multiply, 0.77x for Cholesky, and 0.44x for SVD. Currently cloud providers charge a 2x premium in CPU-core-seconds pricing for their serverless offerings, so numpywren is cheaper for the SVD algorithm. With increased efficiencies and greater competition, we anticipate the premium ratio will fall further, expanding the cost-effectiveness of numpywren.

While these results are promising, using serverless computing for linear algebra has several limitations. First, numpywren is only able to compete with dedicated implementations for large problem sizes (generally larger than $256K \times 256K$ dense matrix operations); the high latency of the external storage makes the system uncompetitive for smaller problem instances.  A more serious limitation is that existing cloud functions platforms cannot efficiently implement the broadcast communication pattern (Figure~\ref{fig:std_comm_patterns}), employed by several popular algorithms such as QR decomposition. This inefficiency  arises because each cloud function provides very few cores (typically no more than 1 or 2), and because the application has no control function placement.   

As an example of communication inefficiency, consider $N$ large machines or VMs, each of them having $K$ cores. A broadcast operation need only send $N$ copies of the data over the network, one to each machine, where all cores can share the same copy. If we assume instead an equivalent deployment using cloud functions, where each function has a single core on a separate computer, then we need to send $N \times K$ data copies over the network to reach the same number of cores. As such the network traffic in this case is $K$ times higher, a non-trivial overhead. Figure~\ref{fig:std_comm_patterns}(a) illustrates this example for $N = K = 2$, by assuming each node has 2 cores, and each function uses a single core, and hence can run two functions.

\subsection{Cirrus: Machine learning training}
Machine learning researchers have traditionally used clusters of VMs for different tasks in ML workflows such as preprocessing, model training, and hyperparameter tuning. One challenge with this approach is that different stages of a pipeline can require significantly different amounts of resources. As with  linear algebra algorithms, a fixed cluster size will either lead to severe underutilization or to severe slowdown. Serverless computing can address this challenge by scaling each stage independently scale to meet its resource demands. Further, it frees developers from managing these servers.

Recent work on the Cirrus project offers promise for ML training pipelines on serverless computing. Cirrus leverages three main observations. First, existing serverless offerings provide linear scalability of compute (cloud functions) and storage throughput (e.g., to S3) up to thousands of cores. Second, caching and prefetching training data can saturate the CPU. Third, a relatively small amount of high-performance storage can significantly improve the performance of these pipelines, and, in particular, of training. In the case of Cirrus, we provide this high-performance storage using  a few VM instances to implement an in-memory parameter server \cite{li2014scaling}.

Using serverless computing for ML workloads is still challenging. First, the gradient needs to be broadcast to every cloud function. As noted in the previous section and as Figure~\ref{fig:std_comm_patterns} illustrates, using cloud functions incurs a much higher communication overhead versus large VM instances. Second, the parameter server must handle asynchronous fine grain updates. This can significantly strain its network connection, both on account of bandwidth and number of packets.

We've evaluated Cirrus against three VM-based ML frameworks: Tensorflow, Bosen and Spark. On a Sparse Logistic Regression workload Cirrus with 10 lambda workers converges 3x faster than Tensorflow (1 x m5d.4xlarge, 32 cores) and 5x faster than Bosen (2 x m5d.4xlarge, 16 cores). On a Collaborative filtering workload Cirrus converges in less time to a lower loss (RMSE of 0.83 vs. 0.85).  While Cirrus can outperform serverful in terms of training completion time, it does not outperform on cost, which may be up to 7x higher.

\subsection{Serverless SQLite:  Databases}
Stateful workloads, such as databases, are particularly challenging for serverless computing. At first glance, these services embody the antithesis of the stateless nature of serverless computing. While cloud providers offer many managed database services with some elasticity \cite{dynamodb, clouddatastore, cloudspanner, cosmosdb, aurora}, an intriguing question is whether a third party could implement a serverless database directly on top of a serverless computing platform.

A strawman solution would be to run common transactional databases, such as PostgreSQL, Oracle, or MySQL inside serverless functions. However, that immediately runs into a number of challenges. First, serverless computing has no built-in persistent storage, so we need to use a  remote persistent store, which introduces large latency. Second, these databases assume connection-oriented protocols, i.e., databases run as services accepting connections from clients. This assumption conflicts with existing  cloud functions that are running behind network address translators, and thus don't support incoming connections. Finally, while many high performance databases rely on shared memory \cite{hellerstein2007architecture}, cloud functions run in isolation so cannot share memory. Shared-nothing distributed databases \cite{corbett2013spanner, cockroachdb, NuoDB} do not require shared memory, but they expect nodes to be directly addressable on the network and they expect cluster membership to change only slowly. All these issues pose significant challenges to running traditional database software atop of serverless computing, or to implementing equivalent functionality.

Despite these challenges, we did succeed in running the SQLite embedded database on a serverless computing platform. SQLite runs as an application library, so it doesn't need to support inbound network connections. Also, SQLite doesn't require shared memory and instead relies on access to a fast, cached, shared file system. Our approach to providing such a file system in a serverless environment, is to interpose an in-memory transactional caching buffering layer between SQLite and the cloud provider's network file system (e.g., AWS EFS or Azure Files). By maintaining a change log in a hidden file, we provide both transactional isolation and effective caching for shared file system access. This scales as does serverless compute, supporting hundreds of concurrent functions and sub-millisecond latencies to achieve over 10 million tpmC (transactions per minute) on a modified read-only TPC-C benchmark \cite{tpcc}, performance comparable to that which commercial RDBMS systems report on the unmodified benchmark. Whereas these unmodified benchmarks comprise about 70\% writes, our scalability with writes only reaches a tpmC of roughly 100 due to reliance on database-level locking.

Our conclusion is that most database-like applications will continue to be provided as BaaS, unless the application naturally makes writes to the database rare.

\end{document}